\documentclass[letterpaper,12pt]{article}   
\usepackage{osajnl2}

\begin{document}

\title{Error propagation in polarimetric demodulation}


\author{A. Asensio Ramos$^{1,*}$ and M. Collados$^{1}$}
\address{$^1$Instituto de Astrof\'{\i}sica de Canarias, 38200, La Laguna, Spain}
\address{$^*$Corresponding author: aasensio@iac.es}

\begin{abstract}
The polarization analysis of the light is typically carried out using modulation schemes. The light of unknown
polarization state is passed through a set of known modulation optics and a detector is used to measure
the total intensity passing the system. The modulation optics is modified several times and, with the
aid of such several measurements, the unknown polarization state of the light can be inferred. How to 
find the optimal demodulation process has been investigated in the past. However, since the modulation matrix 
has to be measured for a given instrument and the optical elements can present problems of repeatability, some 
uncertainty is present in the elements of the modulation matrix and/or covariances between
these elements. We analyze in
detail this issue, presenting analytical formulae for calculating the covariance matrix 
produced by the propagation of such uncertainties
on the demodulation matrix, on the inferred Stokes parameters and on the efficiency of the modulation
process. We demonstrate that, even if the covariance matrix of the modulation matrix
is diagonal, the covariance matrix of the demodulation matrix is, in general, non-diagonal because matrix
inversion is a nonlinear operation. This propagates through the demodulation process and induces correlations
on the inferred Stokes parameters.
\end{abstract}


\maketitle 

\section{Introduction}
The majority of the information that we have obtained during the last years about the magnetism 
of the Sun and other astrophysical objects is based on the analysis of the polarization of the light. The
polarization state of a light beam is typically described, using the Stokes formalism, by the 
so-called Stokes vector:
\begin{equation}
\mathbf{S} = (I,Q,U,V)^T,
\label{eq:stokes_vector}
\end{equation}
where $I$ refers to the total intensity of the beam, $V$ describes its circular polarization properties while
$Q$ and $U$ are used to described linear polarization. When the light beam passes through optical elements 
or, in general, any medium that modifies its polarization properties, the emergent Stokes vector 
can be related to the input one by the following linear relation:
\begin{equation}
\mathbf{S}^\mathrm{out} = \mathbf{M} \mathbf{S}^\mathrm{in}.
\end{equation}
The $4 \times 4$ matrix, $\mathbf{M}$, is the so-called Mueller matrix and it can be used to unambiguously 
represent any optically passive medium. It is important to note that these matrices have to obey certain 
properties so that they have physical meaning \cite{givens_kostinski,landi_deltoro98}.

The analysis of polarization of a given beam is usually carried out with the aid of modulation schemes. This is
a requisite in the short-wavelength (optical) domain because no detectors that present the required polarization
sensitivity are still available. Furthermore, the majority of the detectors that are used in the
optical are only sensitive to the total 
amount of light or, in other words, to the total intensity given by Stokes $I$. This
is not the case in the long-wavelength (microwave) domain, where it is relatively easy to build 
detectors that analyze directly the polarization
properties of the light beam. Modulation schemes have been built to overcome the difficulties when 
measuring the polarization state of light beams in the short-wavelength domain. Any modulation scheme consists 
of a train of optical devices that produce a known
modification of the input beam so that the observed intensity in the detector is a linear combination of the
elements of the input Stokes vector. Carrying out several of these measurements, each one with a different combination of 
optical devices, it is possible to \emph{infer} the input Stokes vector using a demodulation procedure. 
Assuming that the Mueller matrix of the
$j$-th combination of optical devices is given by $\mathbf{M}_j$, the detected intensity in each case is given by:
\begin{equation}
I^\mathrm{out}_j = [M_j]_{00}S^\mathrm{in}_0+[M_j]_{01}S^\mathrm{in}_1+[M_j]_{02}S^\mathrm{in}_2+[M_j]_{03}S^\mathrm{in}_3,
\end{equation}
where we have used the standard notation $\mathbf{S} = (S_0,S_1,S_2,S_3)^T$ for the Stokes vector, which 
will be used extensively in the rest of the paper.
By putting several measurements together, the modulation scheme can be expressed as the following linear system:
\begin{equation}
\mathbf{I^\mathrm{out}} = \mathbf{O} \mathbf{S^\mathrm{in}},
\label{eq:modulation}
\end{equation}
where each row of the matrix $\mathbf{O}$ is built with the first rows of the different Mueller matrices of the different
combinations of optical elements used in the modulation scheme. Therefore, the matrix $\mathbf{O}$ has dimensions
$N \times 4$, with $N$ the number of measurements. Note that, neither the vector $\mathbf{I^\mathrm{out}}$ is
a Stokes vector nor the matrix $\mathbf{O}$ is a Mueller matrix. The vector $\mathbf{I^\mathrm{out}}$
consists of different intensity measurements, while the matrix $\mathbf{O}$ does not fulfill, in general, the
conditions described by \cite{givens_kostinski} and \cite{landi_deltoro98} to be considered a Mueller matrix. 

Several different modulation techniques have been developed. One of the most widespread methods is to use a 
temporal modulation with a quarter-wave plate and a linear polarizer, with the fast axis of the quarter-wave
plate being set to preselected angles \cite{povel90,povel95}. This approach has the advantage that the different 
polarizations states
are observed on the same pixel. However, modulation has to be carried out very fast in order not to be
limited by atmospheric fluctuations. Another possibility is to use spatial modulation in which the measurements
are carried out simultaneously by using a beamsplitter instead of a linear polarizer 
\cite{semel93,donati97}. The advantage is that 
atmospheric fluctuations are avoided but potential problems appear because the optical path and the detector 
response are not the same for all the polarization states. The most advanced polarimeters now use a combination
of both approaches in the so-called spatio-temporal modulation \cite{pillet_tip99,lopez_ariste00}.

Starting from Eq. (\ref{eq:modulation}), the input Stokes vector
can be obtained by solving the previous linear system of equations, so that:
\begin{equation}
\mathbf{S^\mathrm{in}} = \mathbf{D} \mathbf{I^\mathrm{out}},
\label{eq:demodulation}
\end{equation}
where the matrix $\mathbf{D}$ is the inverse (or Moore-Penrose pseudoinverse in the general case that more 
than four measurements are carried out to infer the input Stokes vector) of the matrix $\mathbf{O}$ and has
dimensions $4 \times N$. In order to measure the four Stokes parameters, the minimum number 
of measurements is 4. In such a case, the matrix $\mathbf{D}$ is unique and can be calculated 
as the standard inverse of $\mathbf{O}$. However,
it is also possible to carry out more measurements than Stokes parameters. In this case, the linear system
of Eq. (\ref{eq:modulation}) is overdetermined and, in general, no solution that satisfies simultaneously
all the equations exist. 
Recently, it has been shown that it is still possible to choose a solution if we
seek for the one that maximizes the efficiency of the modulation-demodulation scheme \cite{deltoro_collados00}.
Such efficiency for each Stokes parameter, represented with the vector \mbox{\boldmath$\epsilon$},
is obtained from the demodulation matrix \cite{collados99,deltoro_collados00}:
\begin{equation}
\epsilon_i = \left( n \sum_{j=1}^n D_{ij}^2 \right)^{-1/2}.
\label{eq:efficiency}
\end{equation}
In this case, it can be shown that the optimal demodulation matrix is given by the Moore-Penrose 
pseudoinverse \cite{moore20,penrose55}:
\begin{equation}
\mathbf{D} = \left( \mathbf{O}^T \mathbf{O} \right)^{-1} \mathbf{O}^T.
\label{eq:optimal_demodulation}
\end{equation}
The pseudoinverse can also be found with the singular value decomposition of the modulation matrix. Note that,
while $\mathbf{DO}=\mathbf{1}$ in the general case, $\mathbf{OD}=\mathbf{1}$ is only valid when $N=4$ 
($\mathbf{1}$ is the identity matrix).

We present a theoretical calculation of how the error propagates in the demodulation process.
This is an important issue that has been partially treated in the past, either considering
how the uncertainty in the measurement of $\mathbf{I^\mathrm{out}}$
propagates to the demodulated Stokes vector when the modulation matrix is perfectly known
\cite{sabatke00,takakura07} or taking into account uncertainties in the knowledge of the modulation matrix, though
the full covariance matrix is not obtained \cite{hauge78,zallat06}. Many papers also deal with how error is
propagated in non-ideal Mueller matrix polarimeters \cite{nee03,nee06} while
others consider the optimization of such polarimeters \cite{tyo00,tyo02,smith02}.

\section{Error propagation}
In practical situations, the modulation matrix has to be measured once the optics and the modulation scheme have been
defined \cite{tyo02,smith02,nee03,zallat06,takakura07}. As a consequence, one expects the elements of the modulation matrix to have some uncertainty produced
by the measurement procedure. Additionally, uncertainties can also be found when the rotation of the fast axis of
the retarders or of the linear polarizers has some repeatability problems \cite{goldstein_chipman90}. In 
this case, although the average
modulation matrix can be known with great precision, random deviations from the average modulation 
matrix can occur. These random deviations induce a modification on the modulation scheme that directly
affects the inferred polarization properties of the input light beam. For these reasons, it is fundamental
to characterize the error propagation through the demodulation process and how they affect the measured 
Stokes profiles. In general, since the matrix inversion (or pseudoinverse) is a nonlinear process, one should 
expect to find a full covariance matrix for the elements of the matrix $\mathbf{D}$, with non-zero non-diagonal 
elements. These non-diagonal elements account for the statistical correlations that appear
between elements of the demodulation matrix after the inversion process.
More importantly, a full covariance matrix has to be expected even in the case 
that the elements of the modulation matrix are not correlated (i.e., 
$\mathrm{cov}(O_{ij},O_{kl}) \propto \delta_{ik} \delta_{jl}$, with $\delta_{ik}$ the
Kronecker delta). Note that a diagonal covariance matrix for the matrix $\mathbf{O}$ comes out 
when the measurements of the elements of the modulation matrix are statistically independent, something
that can be assumed in some cases if the measurements of the elements of the matrix
are carried out with proper calibration. However, this is an issue that has to be carefully 
analyzed for each case.

\subsection{Analytical approach}
The full covariance matrix for the demodulation matrix can be obtained analytically following the standard
error propagation formulae. The starting point is to consider that elements of the demodulation matrix are
known nonlinear functions of the elements of the modulation matrix:
\begin{equation}
D_{\alpha \beta} = D_{\alpha \beta} (O_{ij}).
\end{equation}
The most general error propagation formula for such operation reads \cite{hahn_shapiro67}: 
\begin{equation}
\mathrm{cov}(D_{\alpha \beta},D_{ab}) = \sum_{ijkl} \frac{\partial D_{\alpha \beta}}{\partial O_{ij}} 
\frac{\partial D_{ab}}{\partial O_{kl}}
\mathrm{cov}(O_{ij},O_{kl}).
\label{eq:covarianceD_general}
\end{equation}
This covariance matrix gives information about the variances of the elements of the demodulation matrix
(diagonal elements), as well as covariances between those elements (non-diagonal elements). Note that, for a
$4 \times N$ matrix, the covariance matrix is $4N \times 4N$. If the covariance matrix of the modulation matrix
is diagonal (i.e., $\mathrm{cov}(O_{ij},O_{kl})=\sigma(\mathbf{O})^2_{ij} \delta_{ik} \delta_{jl}$, with $\delta_{ik}$ the
Kronecker delta), the previous equation
simplifies to:
\begin{equation}
\mathrm{cov}(D_{\alpha \beta},D_{ab}) = \sum_{ij} \frac{\partial D_{\alpha \beta}}{\partial O_{ij}} 
\frac{\partial D_{ab}}{\partial O_{ij}}
\sigma(\mathbf{O})^2_{ij}.
\label{eq:covarianceD_diagonal}
\end{equation}
As already presented above, this turns out to be a good approximation if the elements of the modulation matrix
are carefully measured. However, our approach is general and can cope with non-diagonal covariance matrices
as will be shown below. 

The derivatives that appear in Eqs. (\ref{eq:covarianceD_general}) and 
(\ref{eq:covarianceD_diagonal}) can be calculated, in the most general case, starting from 
Eq. (\ref{eq:optimal_demodulation}). However, we distinguish the cases of a square modulation matrix
from the general case, because the analytical procedure can be largely simplified. In any case, we will
show that the general expressions are equivalent to those of the particular case of a square matrix $\mathbf{O}$.

\subsubsection{Square modulation matrix}
Although it is possible to calculate directly the derivatives from Eq. (\ref{eq:optimal_demodulation}),
it is more advantageous to use the fact that the matrix product between $\mathbf{O}$ and $\mathbf{D}$ commute, so 
that $\mathbf{D}\mathbf{O}=\mathbf{O}\mathbf{D}=\mathbf{1}$. In this case, it is easy to show 
that \cite{lefebvre00}:
\begin{equation}
\frac{\partial D_{\alpha \beta}}{\partial O_{ij}} = -D_{\alpha i} D_{j \beta}.
\label{eq:derivative_square}
\end{equation}
Therefore, the full covariance matrix of the matrix elements of the demodulation
matrix can be calculated by substituting Eq. (\ref{eq:derivative_square}) into Eq. (\ref{eq:covarianceD_general}):
\begin{equation}
\mathrm{cov}(D_{\alpha \beta},D_{ab}) = \sum_{ijkl} D_{\alpha i} D_{j \beta} D_{a k} D_{l b} 
\mathrm{cov}(O_{ij},O_{kl}).
\end{equation}
In the particular case of a diagonal covariance matrix for $\mathbf{O}$, the previous
equation simplifies to:
\begin{equation}
\mathrm{cov}(D_{\alpha \beta},D_{ab}) = \sum_{ij} D_{\alpha i} D_{j \beta} D_{a i} D_{j b} 
\sigma(\mathbf{O})^2_{ij}.
\label{eq:covariance_square}
\end{equation}

\subsubsection{Non-square modulation matrix}
When $N$ is larger than four, it is not possible to follow the previous
approach because $\mathbf{O}$ and $\mathbf{D}$ do not commute. However, inserting the
intermediate matrix $\mathbf{A}=\mathbf{O}^T \mathbf{O}$, using the fact that
$\mathbf{D}=\mathbf{A}^{-1} \mathbf{O}^T$ and that the matrix $\mathbf{A}$ is always invertible \cite{deltoro_collados00},
the derivative can be expressed, after some algebra, as:
\begin{equation}
\frac{\partial D_{\alpha \beta}}{\partial O_{ij}} = A_{\alpha j}^{-1} \delta_{\beta i} + \sum_k O_{\beta k}
\frac{\partial A_{\alpha k}^{-1}}{\partial O_{ij}}.
\label{eq:deriv_D_O}
\end{equation}
The derivative of the elements of the $\mathbf{A}^{-1}$ matrix with respect to the elements of $\mathbf{O}$
can be calculated using the chain rule:
\begin{equation}
\frac{\partial A_{\alpha k}^{-1}}{\partial O_{ij}} = \sum_{mn}
\frac{\partial A_{\alpha k}^{-1}}{\partial A_{mn}} \frac{\partial A_{mn}}{\partial O_{ij}}.
\label{eq:der_Ainv}
\end{equation}
The definition of the matrix $\mathbf{A}$ allows us to obtain the following derivative easily:
\begin{equation}
\frac{\partial A_{mn}}{\partial O_{ij}} = O_{in} \delta_{mj} + O_{jm} \delta_{nj}.
\end{equation}
The derivative of the elements of $\mathbf{A}^{-1}$ with respect to the elements of $\mathbf{O}$
can also be calculated easily from Eq. (\ref{eq:derivative_square}) because $\mathbf{A}$ is a
square matrix, that commutes with its inverse. Therefore:
\begin{equation}
\frac{\partial A^{-1}_{\alpha \beta}}{\partial O_{ij}} = -A^{-1}_{\alpha i} A^{-1}_{j \beta}.
\label{eq:der_Ainvfinal}
\end{equation}
Substituting Eq. (\ref{eq:der_Ainvfinal}) into Eq. (\ref{eq:deriv_D_O}) and after some algebra,
we end up with the following expression for the derivative:
\begin{equation}
\frac{\partial D_{\alpha \beta}}{\partial O_{ij}} = A_{\alpha j}^{-1} \delta_{\beta i} - 
A_{\alpha j}^{-1} \sum_n D_{n \beta} O_{in} - D_{\alpha i} D_{j \beta}.
\label{eq:derivative_nonsquare}
\end{equation}
This expression reduces to Eq. (\ref{eq:derivative_square}) if $\mathbf{O}$ is square because 
the product of $\mathbf{D}$ and $\mathbf{O}$ commutes, so that:
\begin{equation}
\sum_n D_{n \beta} O_{in} = \sum_n O_{in} D_{n \beta} = \delta_{i \beta},
\end{equation}
After substitution in Eq. (\ref{eq:derivative_nonsquare}), we recover Eq. (\ref{eq:derivative_square}). 

The final expression for the
covariance of the demodulation matrix is obtained by plugging Eq. (\ref{eq:derivative_nonsquare}) into
Eq. (\ref{eq:covarianceD_general}) [or Eq. (\ref{eq:covarianceD_diagonal})]:
\begin{eqnarray}
\mathrm{cov}(D_{\alpha \beta},D_{ab}) &=& 
\sum_{ijkl} \mathrm{cov}(O_{ij},O_{kl}) \nonumber \\
&\times& \left[ A_{\alpha j}^{-1} \delta_{\beta i} - 
A_{\alpha j}^{-1} \sum_n D_{n \beta} O_{in} - D_{\alpha i} D_{j \beta} \right] \nonumber \\
&\times& \left[ A_{ak}^{-1} \delta_{bk} - 
A_{ak}^{-1} \sum_m D_{mb} O_{km} - D_{ak} D_{lb} \right].
\end{eqnarray}

When both the modulation matrix $\mathbf{O}$ and the measured intensities
$\mathbf{I^\mathrm{out}}$ present uncertainties, the resulting uncertainty in the demodulated Stokes vector
has contribution coming from both origins. Using standard error propagation formulae applied to Eq. (\ref{eq:demodulation}), 
the covariance matrix can be written as:
\begin{eqnarray}
\mathrm{cov}(S^\mathrm{in}_i,S^\mathrm{in}_j) &=& \sum_{\alpha \beta a b} 
\frac{\partial I^\mathrm{in}_i}{\partial D_{\alpha \beta}} \frac{\partial I^\mathrm{in}_j}{\partial D_{ab}}
\mathrm{cov}(D_{\alpha \beta},D_{ab}) \nonumber \\
&+&\sum_{kl} \frac{\partial S^\mathrm{in}_i}{\partial I^\mathrm{out}_k} 
\frac{\partial S^\mathrm{in}_j}{\partial I^\mathrm{out}_l}
\mathrm{cov}(I^\mathrm{out}_k,I^\mathrm{out}_l).
\end{eqnarray}
The first contribution takes into account the uncertainty in the knowledge of the modulation matrix while
the second contribution takes into account the uncertainty in the measurement of the intensities arriving
to the detector \cite{sabatke00,dereniak00}. In the field of optimization of polarimeters, a suitable norm
of this very last term (often neglecting the non-diagonal covariances in the vector $\mathbf{I}^\mathrm{out}$ that we
include here) is the chosen one to measure the efficiency of a polarimeter \cite{ambirajan95,tyo00,sabatke00,dereniak00}.
Particularizing to our problem and calculating the partial derivatives, we end up with:
\begin{eqnarray}
\mathrm{cov}(S^\mathrm{in}_i,S^\mathrm{in}_j) &=& \sum_{\alpha \beta} I^\mathrm{out}_\alpha I^\mathrm{out}_\beta
\mathrm{cov}(D_{i \alpha},D_{j \beta}) \nonumber \\
&+&\sum_{kl} D_{ik} D_{jl} \mathrm{cov}(I^\mathrm{out}_k,I^\mathrm{out}_l).
\label{eq:covariance_intensity}
\end{eqnarray}
The covariance matrix $\mathrm{cov}(\mathbf{I}^\mathrm{out})$ is
typically assumed to be diagonal, since no statistical dependence is assumed between consecutive intensity
measurements in the detector, so that $\mathrm{cov}(I^\mathrm{out}_k,I^\mathrm{out}_l) \propto \delta_{kl}$. 
Since this is not generally the case for the quantities $\mathrm{cov}(D_{i \alpha},D_{j \beta})$, 
a non-zero (in general) covariance between different Stokes profiles appears as a consequence of some degree of 
correlation in the demodulation process.

In principle, it is possible to diagonalize the covariance matrix given by Eq. (\ref{eq:covariance_intensity})
to obtain its principal components. Such principal components define the directions along which the correlation 
between different Stokes parameters is minimized. Since the principal components define a new reference system, it 
is possible to rotate the original reference system in which the four-dimensional Stokes vector is defined. This 
rotation minimizes in a statistical sense the cross-talk. This could be of interest when synthetic calculations
have to be compared with polarimetric observations. In this case, in order to minimize the effect of
cross-talk, it could be advantageous to compare the observations with projections along the eigenvectors 
of the synthetic calculations.

A quantity that is also affected by uncertainties in the knowledge of $\mathbf{O}$ is the efficiency
defined in Eq. (\ref{eq:efficiency}). Applying the error propagation formula, we get:
\begin{equation}
\mathrm{cov}(\epsilon_\alpha,\epsilon_\beta) = \sum_{ijkl} \frac{\partial \epsilon_\alpha}{\partial O_{ij}} 
\frac{\partial \epsilon_\beta}{\partial O_{kl}}
\mathrm{cov}(O_{ij},O_{kl}).
\label{eq:covariance_efficiency}
\end{equation}
If we assume that the covariance matrix
of the elements of the modulation matrix is diagonal, we get:
\begin{equation}
\mathrm{cov}(\epsilon_\alpha,\epsilon_\beta) = \sum_{ij} \frac{\partial \epsilon_\alpha}{\partial O_{ij}} 
\frac{\partial \epsilon_\beta}{\partial O_{ij}}
\sigma(\mathbf{O})^2_{ij}.
\label{eq:covariance_efficiency_diagonal}
\end{equation}
The derivatives can be calculated using the definition of the efficiency and the chain rule:
\begin{equation}
\frac{\partial \epsilon_\alpha}{\partial O_{ij}} = \sum_{pq} \frac{\partial \epsilon_\alpha}{\partial D_{pq}}
\frac{\partial D_{pq}}{\partial O_{ij}},
\label{eq:derivative_epsilon}
\end{equation}
where
\begin{equation}
\frac{\partial \epsilon_\alpha}{\partial D_{pq}} = - D_{\alpha q} \delta_{p \alpha} 
n^{-1/2} \left( \sum_l D_{\alpha l}^2 \right)^{-3/2}.
\end{equation}
Plugging this expression into Eq. (\ref{eq:derivative_epsilon}), we can simplify it to read:
\begin{equation}
\frac{\partial \epsilon_\alpha}{\partial O_{ij}} = -n^{-1/2} \left( \sum_l D_{\alpha l}^2 \right)^{-3/2}
\sum_{q} D_{\alpha q} \frac{\partial D_{\alpha q}}{\partial O_{ij}}.
\end{equation}
The derivative $\partial D_{\alpha q}/\partial O_{ij}$ is obtained from Eq. (\ref{eq:derivative_nonsquare}) 
in the general case and from Eq. (\ref{eq:derivative_square}) in the case of a square modulation matrix.

According to the previous results, when the covariance matrix of the modulation matrix is diagonal and the
variance is the same for all the elements, all the covariance matrices calculated in this section are 
proportional to the variance of the elements of $\mathbf{O}$. As a consequence, reducing an order of magnitude
the uncertainty in the elements of the modulation matrix reduces an order of magnitude the uncertainty in the
demodulation matrix and in the inferred Stokes vector.

\subsection{Monte Carlo approach}
The previous analytical approach assumes that the error propagates normally. This is typically a good 
approximation when the modulation matrix is far from singular. In other words, the rows of the modulation matrix
have to be as linearly independent as possible. However, since the inversion is a
nonlinear process, it would be possible to obtain distributions in the demodulation matrix (and as a consequence,
in the inferred Stokes parameters) that are far from Gaussians. In order to verify this issue and also
to test that the previous analytical approach gives the correct answer, we have carried out a Monte Carlo
experiment. For simplicity, we assume that the measurement uncertainty is the same for all the matrix elements 
of the modulation matrix and that it is characterized by the variance $\sigma^2$, so that the full variance matrix 
is given by:
\begin{equation}
\sigma(\mathbf{O})^2_{ij} = \sigma^2.
\end{equation}
This selection makes the presentation of the results easier, but the previous approach is general and can
cope with non-diagonal covariance matrices. The Monte Carlo experiment, given an initial 
modulation matrix, generates $N$ instances of such a matrix with added uncertainties:
\begin{equation}
\mathbf{\hat{O}}_i = \mathbf{O} + \gamma_i \mbox{\boldmath$\sigma$}(\mathbf{O}),
\end{equation}
where $\gamma_i$ is a normally distributed random constant with zero mean and unit variance. For each
modulation matrix, we obtain the optimal demodulation matrix following Eq. (\ref{eq:optimal_demodulation}).
Finally, the statistical properties of the demodulation matrix are characterized by the full covariance
matrix, given by:
\begin{equation}
\mathrm{cov}(D_{ij},D_{kl}) = \mathrm{E}[D_{ij} D_{kl}] - \mathrm{E}[D_{ij}] \mathrm{E}[D_{kl}],
\end{equation}
where $\mathrm{E}[x]$ stands for the expected value of the $x$ random variable. In such a Monte Carlo approach,
we estimate it using the sample mean, so that $\mathrm{E}[x] = \langle x \rangle$.

\section{Illustrative examples}
The previous analytical approach is applied to several modulation matrices obtained from the literature
and that represent different typical modulation schemes. We analyze the uncertainties in the
demodulation matrix and in the inferred Stokes parameters using the error propagation formulation presented in
the previous section. For illustration purposes, the results are also accompanied in some cases with plots 
obtained using the Monte Carlo approach. One of the reasons for that is to demonstrate whether the resulting 
distributions can be correctly 
assumed to be Gaussians. The Monte Carlo results have been obtained 
using $N=60000$. We assume that the variance that
characterizes the uncertainty in the measurement of the modulation matrix is $\sigma^2=6.25 \times 10^{-6}$
(the standard deviation is equal to $\sigma=2.5 \times 10^{-3}$), which is a reasonable value for a
standard calibration of the modulation. 

\subsection{Square modulation matrix}
\begin{figure*}
\includegraphics[width=1\textwidth]{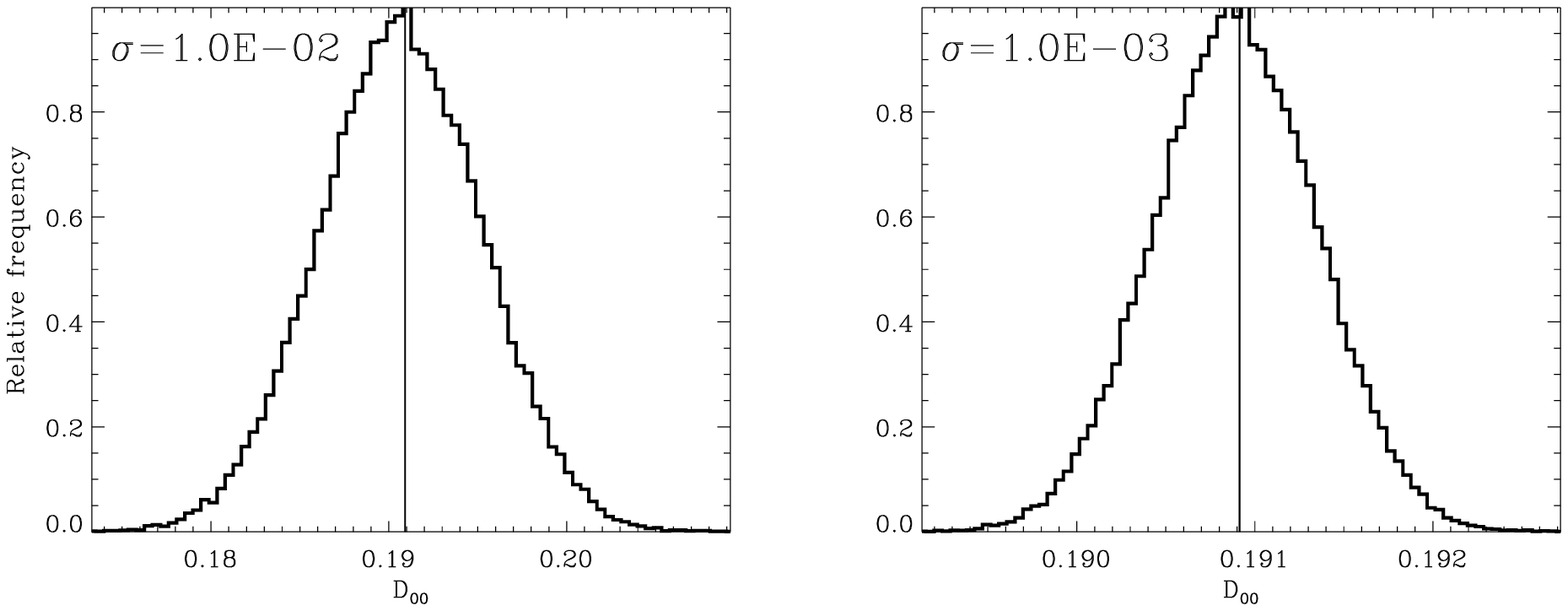}
\caption{Distribution of the $D_{00}$ matrix element of the TIP demodulation matrix when the modulation matrix
elements are known with the uncertainty indicated in each panel. Note that the distribution is centered
at the value obtained from applying Eq. (\ref{eq:optimal_demodulation}) but a dispersion is present around
this value.}
\label{fig:D00_tip}
\end{figure*}

The first example is representative of a square modulation matrix in which four measurements are
carried out to infer the four Stokes parameters. The matrix is 
that used by the Tenerife Infrared Polarimeter (TIP; \cite{martinez_pillet99}) as presented in \cite{deltoro_collados00}. 
Although this matrix is not
the experimental one for TIP, its presents the desired structure and can be used to gain some insight on the 
properties of the error propagation. The chosen modulation matrix is:
\begin{equation}
\mathbf{O}_\mathrm{TIP} = \left[ \begin{array}{cccc}
1 & 0.47 & -0.68 & 0.48 \\
1 & -0.91 & -0.19 & -0.13 \\
1 & -0.11 & 0.57 & 0.72 \\
1 & 0.68 & 0.27 & -0.58
\end{array} \right]
\end{equation}
The optimal demodulation matrix, obtained from the application of Eq. (\ref{eq:optimal_demodulation}), is just
the inverse of $\mathbf{O}$:
\begin{equation}
\mathbf{D}_\mathrm{TIP} = \left[ \begin{array}{cccc}
0.19 & 0.31 & 0.18 & 0.31 \\
0.31 & -0.66 & -0.03 & 0.37 \\
-0.72 & -0.24 & 0.67 & 0.29 \\
0.36 & -0.34 & 0.60 & -0.61
\end{array} \right]
\end{equation}
As shown in Fig. \ref{fig:D00_tip}, the distribution of the elements of the demodulation matrix are Gaussians
whose standard deviation can be obtained from the diagonal elements of the matrix defined in Eq. (\ref{eq:covariance_square}):
\begin{equation}
\mbox{\boldmath$\sigma$}(\mathbf{D}_\mathrm{TIP}) = \left[ \begin{array}{cccc}
1.14 &  1.08 &  1.19 &  1.07  \\
1.81 &  1.71 &  1.88 &  1.70  \\
2.34 &  2.22 &  2.43 &  2.20  \\
2.19 &  2.07 &  2.27 &  2.06 
\end{array} \right] \times 10^{-3}.
\end{equation}
Although we only present the diagonal elements of the covariance matrix, note that the covariance matrix
is a full matrix in which all the elements are different from zero.

As a direct consequence of the uncertainties in the modulation matrix, the efficiency of the proposed scheme
presents uncertainties. They can be calculated using 
Eq. (\ref{eq:covariance_efficiency_diagonal}), so that the average value and their standard deviations are:
\begin{eqnarray}
\mbox{\boldmath$\epsilon$}_\mathrm{TIP}&=&(0.969,0.612,0.473,0.506) \nonumber \\
\mbox{\boldmath$\sigma$}(\mbox{\boldmath$\epsilon$}_\mathrm{TIP})&=&(1.38,1.26,1.25,1.27)\times 10^{-3},
\end{eqnarray}
where non-diagonal elements in $\mathrm{cov}(\epsilon_\alpha,\epsilon_\beta)$ are non-zero, but typically
much smaller than the elements in the diagonal.

\begin{figure*}
\includegraphics[width=1\textwidth]{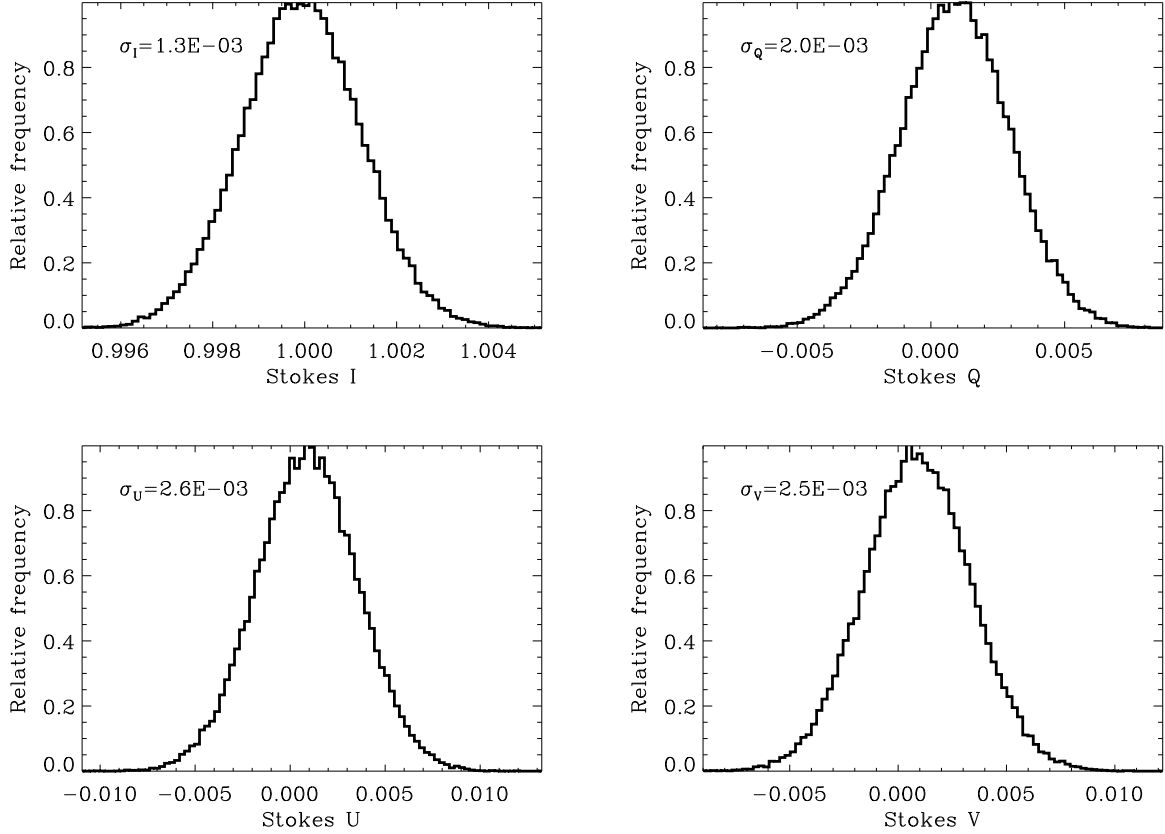}
\caption{Distribution of the inferred Stokes parameters when no uncertainty is assumed in the measurement
of the $\mathbf{I^\mathrm{out}}$ vector but only in the measurement of the TIP modulation matrix, whose
uncertainty is $\sigma=2.5 \times 10^{-3}$. The input Stokes vector was $\mathbf{S^\mathrm{in}}=(1,10^{-3},10^{-3},10^{-3})$.
The distributions are close to Gaussians centered on the original values. The standard deviation of the
inferred value is given in each panel. Note that the errors are roughly similar or slightly smaller than the
uncertainty in the modulation matrix. Note also that correlations between the elements of the
inferred vector are also expected and are produced by the non-diagonal covariance matrix of the demodulation matrix.}
\label{fig:inferred_stokes_tip}
\end{figure*}

\begin{figure*}
\includegraphics[width=1\textwidth]{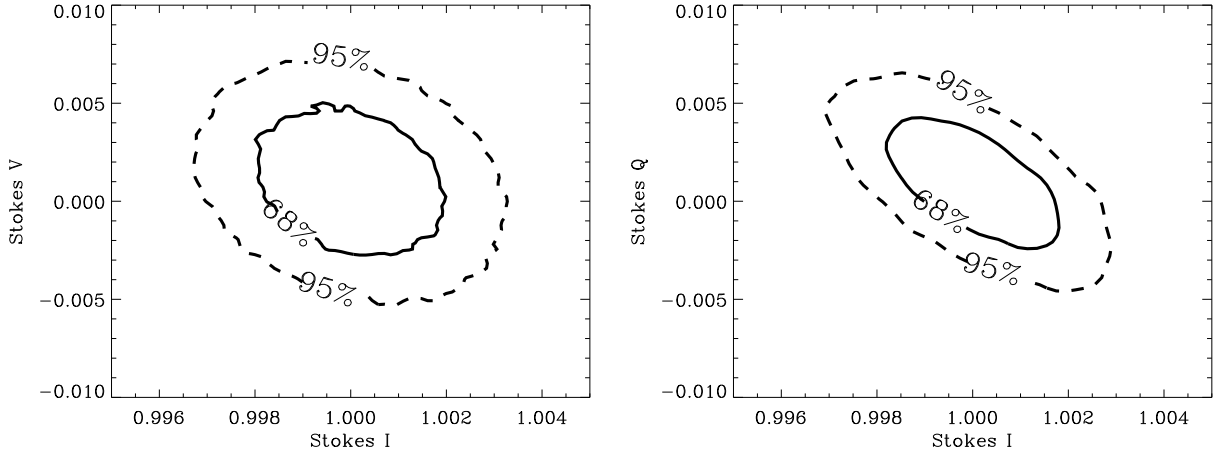}
\caption{Left panel: two-dimensional distribution obtained from the Monte Carlo simulation for $S^\mathrm{in}_0$ and
$S^\mathrm{in}_3$ using TIP's modulation scheme, i.e., the inferred Stokes $I$ and $V$, respectively. The 
plot shows the 
presence of a small correlation between the two results, a direct consequence of the appearance of correlations 
during the inversion process of the modulation matrix. Right panel: the same result but for the ASP case and
for $S^\mathrm{in}_0$ and $S^\mathrm{in}_1$, i.e., the inferred Stokes $I$ and $Q$. Note that the 
correlation between both is even larger, as indicated by the respective
covariance matrices given by Eqs. (\ref{eq:cov_stokes_tip}) and (\ref{eq:cov_stokes_asp}).}
\label{fig:correlationIV}
\end{figure*}

Now we investigate the propagation of uncertainties to the inferred Stokes parameters. To this end,
we choose an initial light beam defined by $\mathbf{S^\mathrm{in}}=(1,10^{-3},10^{-3},10^{-3})$. Such Stokes
vector is representative of what one would observe in relatively low magnetic flux regions of the 
solar surface. The application of Eqs. (\ref{eq:covariance_intensity}) and (\ref{eq:covariance_square}) gives 
the following covariance matrix:
\begin{equation}
\mathrm{cov}(\mathbf{S}^\mathrm{in}) = \left[ \begin{array}{cccc}
 1.66 &  -0.21 &   0.01 &  -0.76\\
-0.21 &   4.17 &   0.14 &   0.59\\
 0.01 &   0.14 &   6.99 &   0.32\\
-0.76 &   0.59 &   0.32 &   6.09
\end{array} \right] \times 10^{-6}
\label{eq:cov_stokes_tip}
\end{equation}
This result has been obtained assuming that there is no uncertainty in the measurement of the intensity arriving to
the detector, so that $\mathrm{cov}(I^\mathrm{out}_k,I^\mathrm{out}_l)=0$. In so doing, we isolate the effect of
the modulation on the inferred Stokes vectors. The results of the Monte Carlo simulation are shown in Fig. 
\ref{fig:inferred_stokes_tip}, demonstrating that the values are quasi-normally distributed around the original value 
with a dispersion that is given in each panel of the plot. Note that the standard deviations are large, almost of the
order of the standard deviation of the precision in the measurement of the modulation matrix. Note also that the
largest value of $\mathrm{cov}(S^\mathrm{in}_i,S^\mathrm{in}_j)$ are on the diagonal of the matrix, but sizable
non-diagonal elements also appear, indicating a certain degree of correlation between the 
inferred Stokes parameters. For instance, the value $\mathrm{cov}(S^\mathrm{in}_0,S^\mathrm{in}_3)=-0.76 \times 10^{-6}$
indicates that when the inferred Stokes $I$ ($S^\mathrm{in}_0$) increases, the inferred 
Stokes $V$ ($S^\mathrm{in}_3$) tends to systematically 
decrease. This can be understood
as a cross-talk between all the Stokes parameters induced by the demodulation process due to the
special structure of the modulation matrix. The two-dimensional
distribution of the inferred Stokes $I$ and $V$ is shown in the left panel of Fig. \ref{fig:correlationIV}. In 
specific cases, part of this cross-talk
can be corrected for easily based on physical arguments. This happens, for instance, when observing the Stokes 
profiles induced by the Zeeman effect in magnetized regions of the solar surface with the aid of 
spectropolarimeters. In such a case, one can assume that Stokes $Q$, $U$ and $V$ are zero away from the
spectral line and carry out a correction of the cross-talk from Stokes $I$ to Stokes $Q$, $U$ and $V$.

The diagonalization of the $\mathrm{cov}(\mathbf{S}^\mathrm{in})$ matrix gives the following eigenvector
matrix ordered in row format:
\begin{equation}
\mathbf{V} = \left[ \begin{array}{cccc}
  0.986 &   0.042 &  -0.012 &   0.160\\
 0.002 &   0.963 &  -0.018 &  -0.268\\
-0.057 &   0.125 &   0.913 &   0.385\\
-0.156 &   0.234 &  -0.408 &   0.868
\end{array} \right].
\label{eq:eigenvectors_tip}
\end{equation}
Each row represents the linear combination of Stokes parameters that one infers due to the 
induced cross-talk contamination. Note that the weight of one of the Stokes parameters is 
larger in each row, but the contamination from the other parameters is still large, a consequence
of the chosen modulation matrix.

\subsection{Non-square modulation matrix}
We present here results for a typical non-square modulation matrix. We have chosen the one belonging to the
Advanced Stokes Polarimeter (ASP; \cite{elmore_ASP92}) as presented in \cite{deltoro_collados00}. As it happens for
the TIP matrix, this is probably not the exact matrix used in the polarimeter but is representative of what happens
in a scheme in which more than four measurements are used to obtain the four Stokes parameters. The modulation
matrix we choose is:
\begin{equation}
\mathbf{O}_\mathrm{ASP} = \left[ \begin{array}{cccc}
1 & 0.77 & 0.41 & -0.36 \\
1 & -0.06 & 0.41 & -0.86 \\
1 & -0.06 & -0.41 & -0.86 \\
1 & 0.77 & -0.41 & -0.36 \\
1 & 0.77 & 0.41 & 0.36 \\
1 & -0.06 & 0.41 & 0.86 \\
1 & -0.06 & -0.41 & 0.86 \\
1 & 0.77 & -0.41 & 0.36 \\
\end{array} \right].
\end{equation}
After calculating the demodulation matrix and its covariance matrix, we can also calculate the diagonal
of the covariance matrix for the efficiency which, transformed into standard deviations, give:
\begin{eqnarray}
\mbox{\boldmath$\epsilon$}_\mathrm{ASP}&=&(0.760,0.415,0.410,0.659) \nonumber \\
\mbox{\boldmath$\sigma$}(\mbox{\boldmath$\epsilon$}_\mathrm{ASP})&=&(1.37,0.94,0.88,0.88)\times 10^{-3},
\end{eqnarray}
and the full covariance
matrix for the inferred Stokes parameters using $\mathbf{S^\mathrm{in}}=(1,10^{-3},10^{-3},10^{-3})$ as the
input Stokes vector:
\begin{equation}
\mathrm{cov}(\mathbf{S}^\mathrm{in}) = \left[ \begin{array}{cccc}
 1.35 &  -1.61 &   0 &   0\\
-1.61 &   4.54 &   0 &   0\\
 0 &   0 &   4.65 &   0\\
 0 &   0 &   0 &   1.80
\end{array} \right] \times 10^{-6}.
\label{eq:cov_stokes_asp}
\end{equation}
Note that the elements of the diagonal of this covariance matrix are smaller than for the case of only four
measurements (except for the case of $\mathrm{cov}(S^\mathrm{in}_1,S^\mathrm{in}_1)$), probably induced 
by the larger number of measurements carried out.
Furthermore, it is important to point out that this modulation scheme induces no correlations 
between $S^\mathrm{in}_2$ (Stokes $U$) and $S^\mathrm{in}_3$ (Stokes $V$) and any other 
Stokes parameter. On the contrary, 
there is a large correlation between Stokes $I$ and Stokes $Q$, 
something that can be clearly seen in the right panel of Fig. \ref{fig:correlationIV}.

The diagonalization of the previous matrix gives the following eigenvector matrix (in row order):
\begin{equation}
\mathbf{V} = \left[ \begin{array}{cccc}
 0.92 &   0.39 &   0 &   0\\
-0.39 &   0.92 &   0 &   0\\
 0 &   0 &   1 &   0\\
 0 &   0 &   0 &   1
\end{array} \right].
\label{eq:eigenvectors_asp}
\end{equation}
The demodulation process has induced a cross-talk between Stokes $I$ and $Q$ which can be represented
by a rotation of $\sim 23^\circ$.

\subsection{Ideal modulation matrix}
It is instructive to present results for the following ideal modulation matrix:
\begin{equation}
\mathbf{O}_\mathrm{IDEAL} = \left[ \begin{array}{cccc}
1 & 1 & 0 & 0 \\
1 & -1 & 0 & 0 \\
1 & 0 & 1 & 0 \\
1 & 0 & -1 & 0 \\
1 & 0 & 0 & 1 \\
1 & 0 & 0 & -1
\end{array} \right].
\end{equation}
The efficiency in this case is the maximum that can be reached, while the diagonal of the covariance matrix 
for the efficiency given as standard deviations is:
\begin{eqnarray}
\mbox{\boldmath$\epsilon$}_\mathrm{IDEAL}&=&(1,\sqrt{3},\sqrt{3},\sqrt{3}) \nonumber \\
\mbox{\boldmath$\sigma$}(\mbox{\boldmath$\epsilon$}_\mathrm{IDEAL})&=&(1.02,1.02,1.02,1.02)\times 10^{-3},
\end{eqnarray}
and the full covariance
matrix for the inferred Stokes parameters using $\mathbf{S^\mathrm{in}}=(1,10^{-3},10^{-3},10^{-3})$ as the
input Stokes vector is:
\begin{equation}
\mathrm{cov}(\mathbf{S}^\mathrm{in}) = \left[ \begin{array}{cccc}
 1.04 &   0 &  0 &  0\\
 0 &   3.13 &   0 &   0\\
 0 &   0 &   3.13 &   0\\
 0 &   0 &   0 &   3.13
\end{array} \right] \times 10^{-6}.
\label{eq:cov_stokes_ideal}
\end{equation}
The covariance matrix is diagonal with equal uncertainties in Stokes $Q$, $U$ and $V$. These values are smaller
than for the ASP example except for Stokes $V$, where the uncertainty for the ideal modulation matrix is slightly
larger. More important is the fact that, since the covariance matrix is diagonal, no correlation is found between
the inferred Stokes parameters, so that no residual cross-talk induced by the modulation is induced in the
demodulation.

\section{Concluding remarks}
We have presented analytical expressions for the calculation of the propagation of errors in the demodulation
process when the modulation matrix is not known with infinite precision. This can happen when the modulation 
system has not been calibrated with enough precision or when the repeatability of the modulation system induces
uncertainties in the modulation scheme. The formulae that we have presented allows polarimeter designers to calculate
the errors in the demodulation matrix, the efficiency of the modulation scheme and the inferred Stokes
parameters. They are simple to calculate and 
require only the knowledge of the modulation matrix, together with its covariance matrix. We 
have pointed out the fact that, in general, since matrix inversion (or Moore-Penrose pseudoinversion) is
a nonlinear operation, non-zero non-diagonal covariances have to be expected in the demodulation matrix even
if such non-diagonal correlations are not present in the modulation matrix. This has the important consequence of 
generating spurious correlations (cross-talk) between the inferred Stokes parameters. We calculate the
induced cross-talk by diagonalizing the covariance matrix. The matrix of eigenvectors represent the
reference system in which the cross-talk is minimized. We have illustrated 
these points with three different modulation matrices representing three different ways of
measuring the polarization state of light beams. Each method has its own advantages and disadvantages.
It is up to the polarimeter designer to choose the modulation scheme depending on the desired precision. 
We hope that the formulae present in this paper are of interest for improving the quality of modulation 
based polarimeters. Routines in FORTRAN 90 and IDL\footnote{\texttt{http://www.ittvis.com/idl}} for the 
calculation of the formulae presented in the 
paper can be obtained after an e-mail request to the authors of this paper.

\section*{Acknowledgments}
Finantial support by the Spanish Ministry of Education and Science through project AYA2007-63881
is gratefully acknowledged.

\bibliographystyle{osajnl}

%
%

\end{document}